\documentclass[12pt]{article}
\usepackage{amssymb,amsmath,graphicx,latexsym}\pdfoutput=1
\begin{document}
\title{\textbf{Far from standard Higgs couplings}} 
\author{B.~Holdom\thanks{bob.holdom@utoronto.ca}\\
\emph{\small Department of Physics, University of Toronto}\\[-1ex]
\emph{\small Toronto ON Canada M5S1A7}}
\date{}
\maketitle
\begin{abstract}
A global fit to LHC data shows a permitted region in the space of couplings of the 126 GeV boson that extends far away from the standard model Higgs couplings. This region is of interest for more natural models of electroweak symmetry breaking. Stronger evidence of vector boson fusion and/or associated production is needed to eliminate this possibility.
\end{abstract}

The Higgs couplings to $W$ and $Z$ are synonymous with the contribution that the Higgs makes to electroweak symmetry breaking. In the standard model this is the only contribution and the couplings are full strength. With more than one Higgs doublet then any particular state, such as the one observed at 126 GeV \cite{:2012gk,:2012gu}, would typically couple with reduced strength to $W$ and $Z$ unless parameters are tuned. Reduced couplings could also suggest that some other more natural mechanism contributes to electroweak symmetry breaking (EWSB). The observation of the 126 GeV boson decaying to $WW$ and $ZZ$ establishes that these couplings exist, but it is still important to firmly establish that these couplings are occurring at full strength.

As we shall explore in this short note, if the gluon fusion production cross section is increased due to new physics and if certain branching ratios are decreased, then various physical observables may be little affected. To the extent that gluon fusion dominates the production, there may be a near degeneracy in the determination of the couplings from the data. Information about other production modes breaks this degeneracy, in particular vector boson fusion and associated production. In the following we shall refer to the 126 GeV boson as the ``Higgs'' even though very non-Higgs-like couplings are considered.

A global fit tailored expressly to display this would-be degeneracy has apparently not been done. To do this we utilize HiggsSignals 1.0 \cite{Bechtle:2013xfa} which incorporates the available data as of April 2013. We first perform a 2 parameter global fit in the space of Higgs couplings with the Higgs mass fixed at 125.7 GeV. Let $g_{gg}$ be a multiplicative factor in the $gg$ coupling (not coupling squared) such that $g_{gg}^{\rm SM}=1$ and similarly $g_X$ is a factor in the $\gamma\gamma$, $VV$ and $\tau\tau$ couplings such that $g_X^{\rm SM}=1$. The $b\overline{b}$, $t\overline{t}$ and other couplings are held at the standard model values. The global fits for ATLAS and CMS results respectively are shown in Fig.~(1).  The combined global fit is presented in Fig.~(2L) where the global minimum is at $g_{gg}=0.87$, $g_X=0.98$ with $\chi^2/{\rm ndf}=26.4/34$.
\begin{figure}[]
\vspace{-5ex}
\centering\includegraphics[scale=.9]{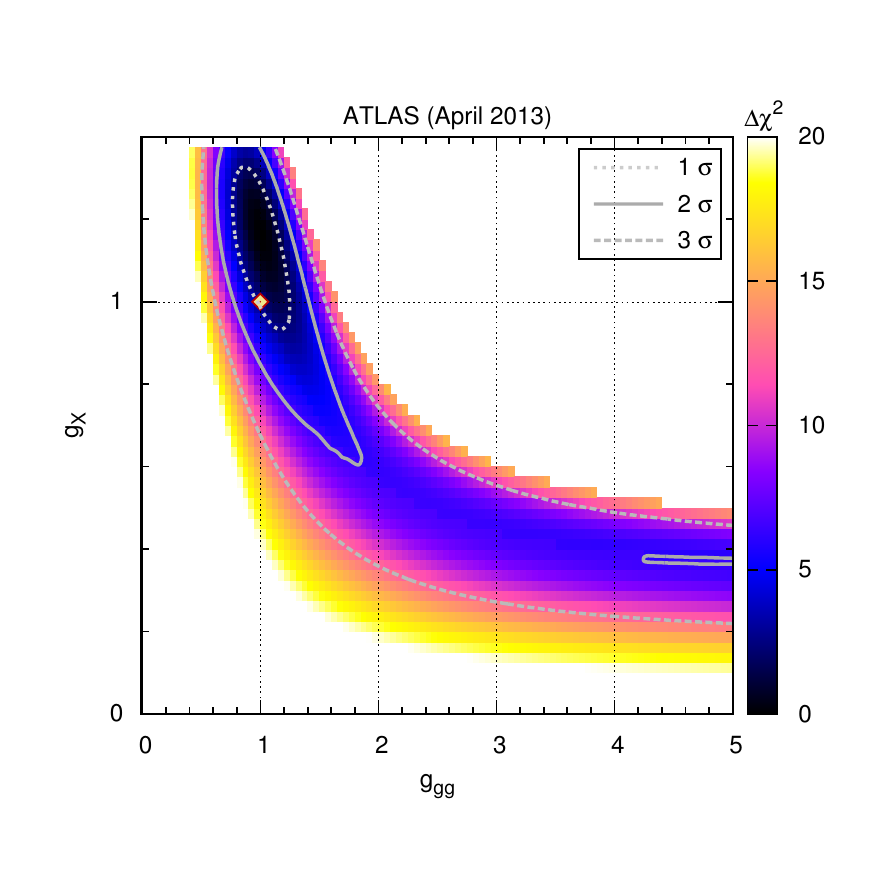}
\centering\includegraphics[scale=.9]{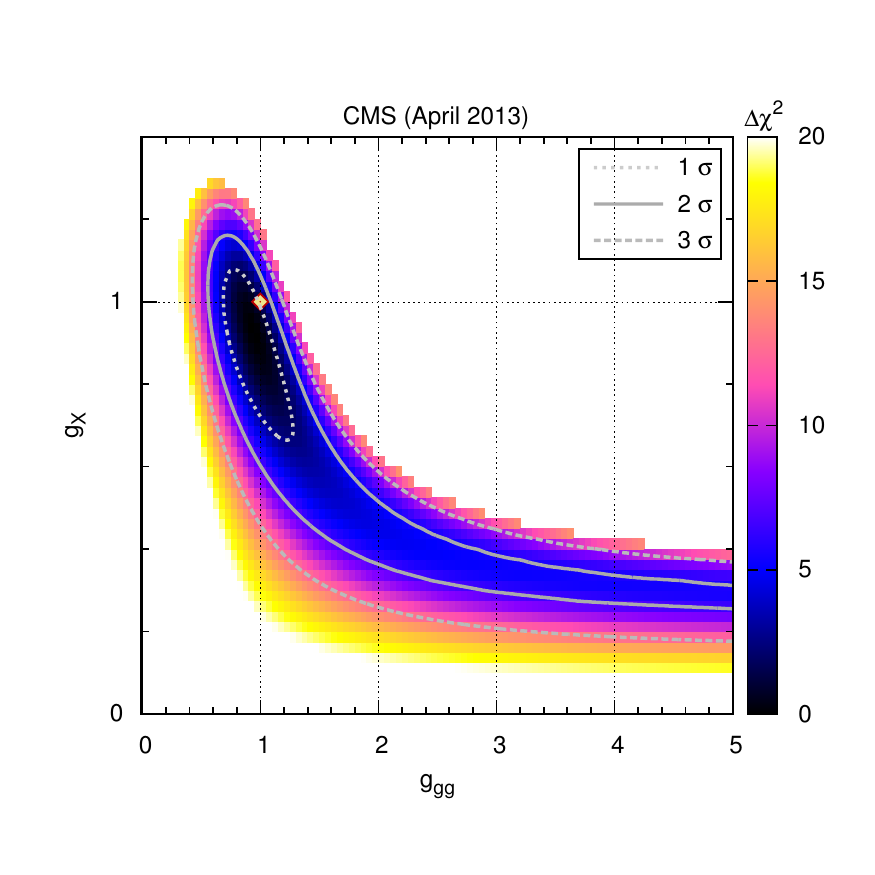}
\vspace{-7ex}
\caption{Two parameter global fits for ATLAS and CMS. $g_{gg}$ is a factor in the $gg$ coupling and $g_X$ is a common factor in the $\gamma\gamma$, $VV$ and $\tau\tau$ couplings. The SM point is marked.}\end{figure}
\begin{figure}[]
\vspace{-5ex}
\centering\includegraphics[scale=.9]{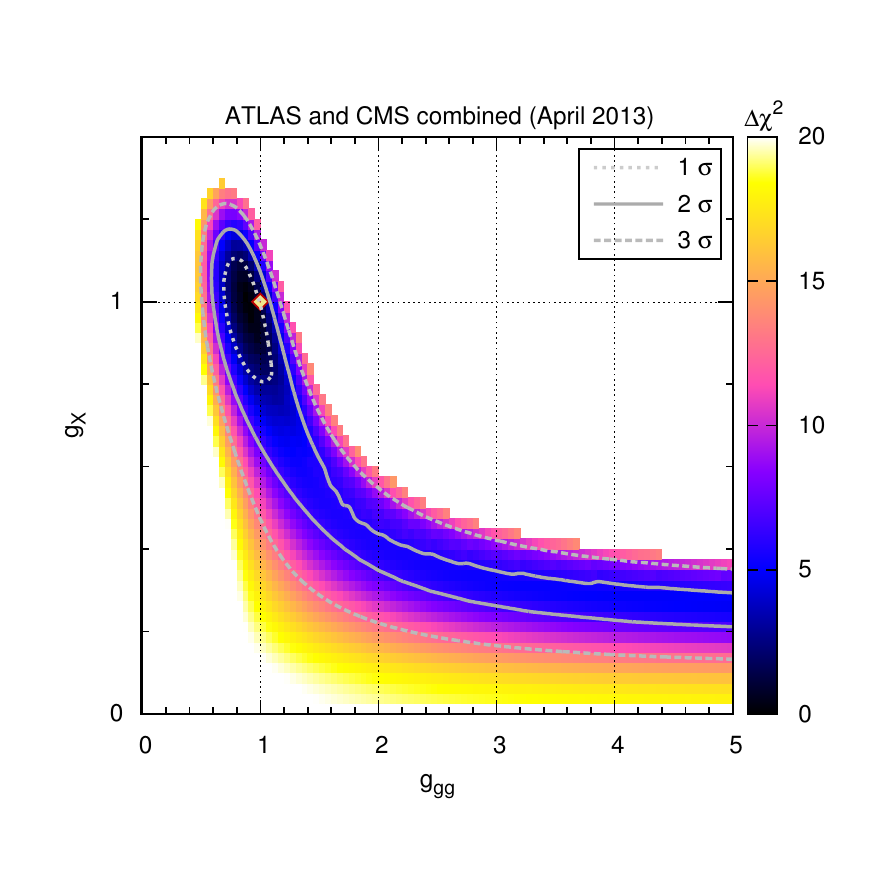}
\centering\includegraphics[scale=.9]{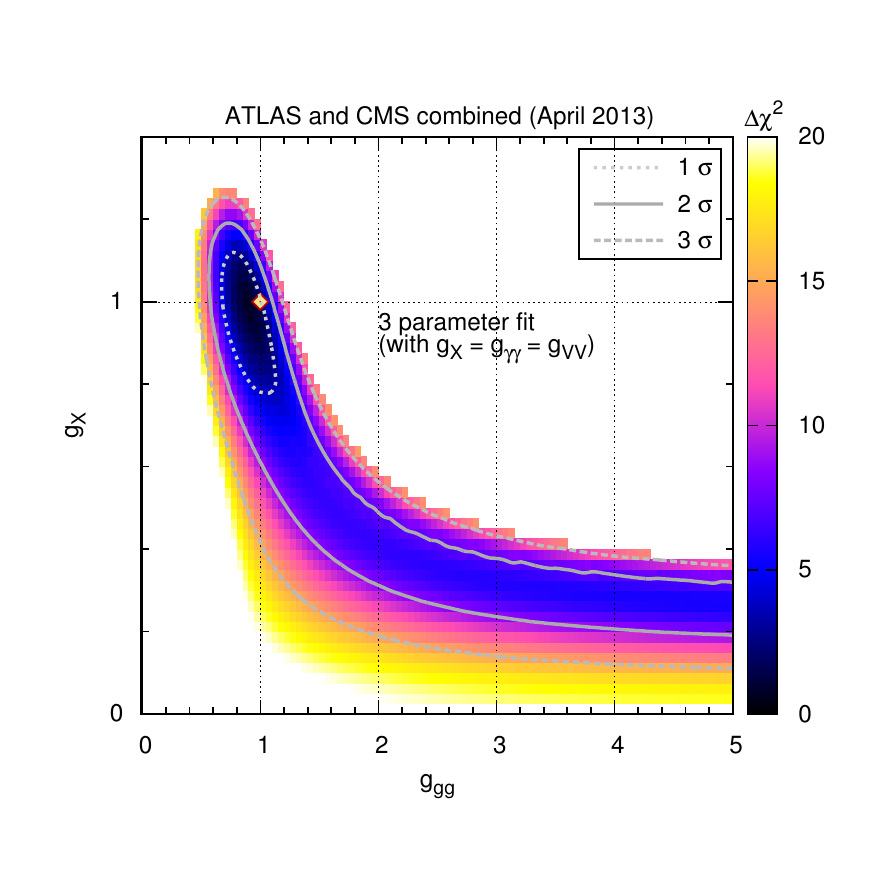}
\vspace{-7ex}
\caption{Left: Same as Fig.~(1) for ATLAS and CMS combined. Right: Three parameter ($g_{gg}$, $g_{\gamma\gamma}$, $g_{VV}$) global fit shown for $g_X= g_{\gamma\gamma}=g_{VV}$.}\end{figure}

In these fits we have kept the relative sizes of the $\gamma\gamma$, $VV$ and $\tau\tau$ couplings as specified by the standard model. Theoretically it makes more sense to relax this assumption and to consider a 3 parameter fit. We choose $g_{gg}$, $g_{\gamma\gamma}$ and $g_{VV}$ as independent and set $g_{\tau\tau}=(g_{\gamma\gamma}+g_{VV})/2$ given less data in the $\tau\tau$ channel. We find the global minimum at $g_{gg}=0.86$, $g_{\gamma\gamma}=1.06$, $g_{VV}=0.96$ with $\chi^2/{\rm ndf}=25.8/33$. The 1, 2, 3 $\sigma$ contours now become surfaces as defined appropriately for a 3 parameter fit. We may consider the intersection of these surfaces with the same plane that was used by the 2 parameter fit. The result is shown in Fig.~(2R) and we see that it is quite similar to Fig.~(2L). In either case the extended regions that are allowed at 2 and 3 $\sigma$ are due to the would-be degeneracy and the preference to the region within the 1 $\sigma$ contour is due to what evidence there is for vector boson fusion and associated production at the LHC.
\begin{figure}[]
\vspace{-6ex}
\centering\includegraphics[scale=.9]{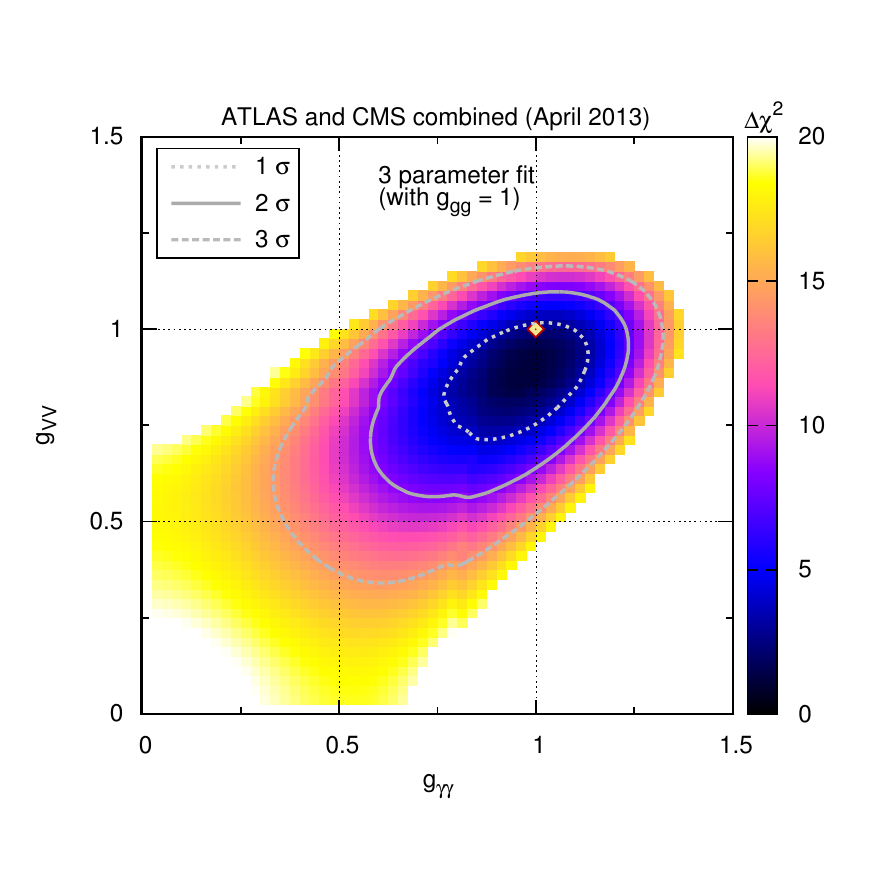}
\centering\includegraphics[scale=.9]{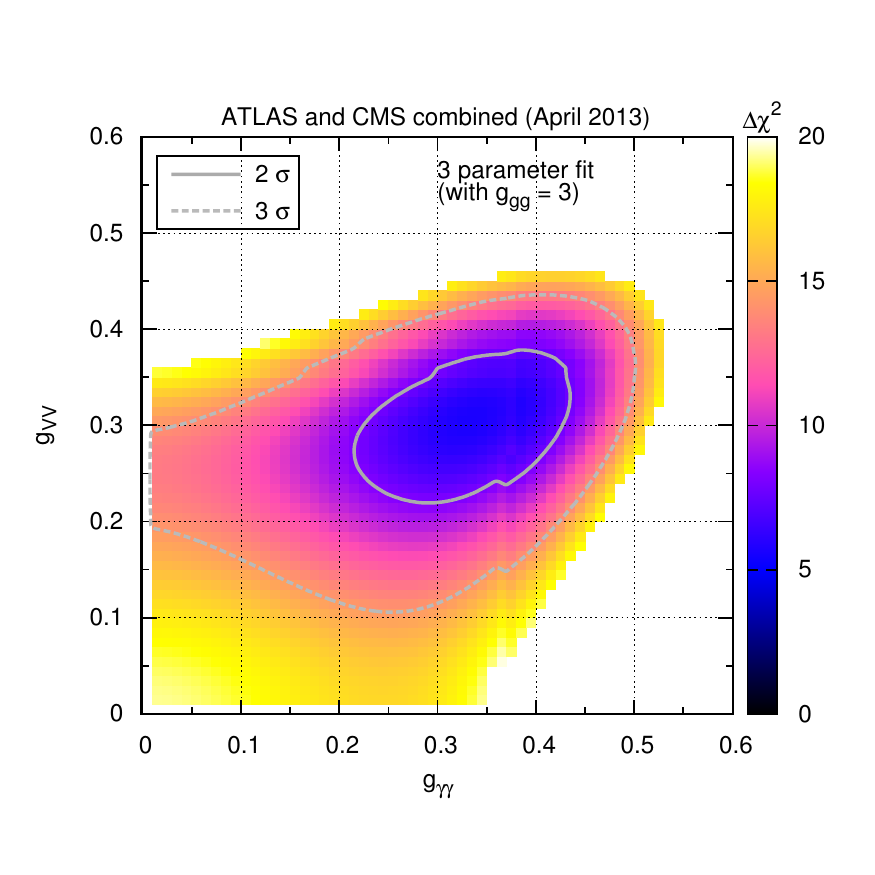}
\vspace{-7ex}
\caption{Three parameter  ($g_{gg}$, $g_{\gamma\gamma}$, $g_{VV}$) global fit shown for $g_{gg}=1$ (left) and $g_{gg}=3$ (right). The left plot includes the SM point.}\end{figure}
\begin{figure}[]
\vspace{-5ex}
\centering\includegraphics[scale=.9]{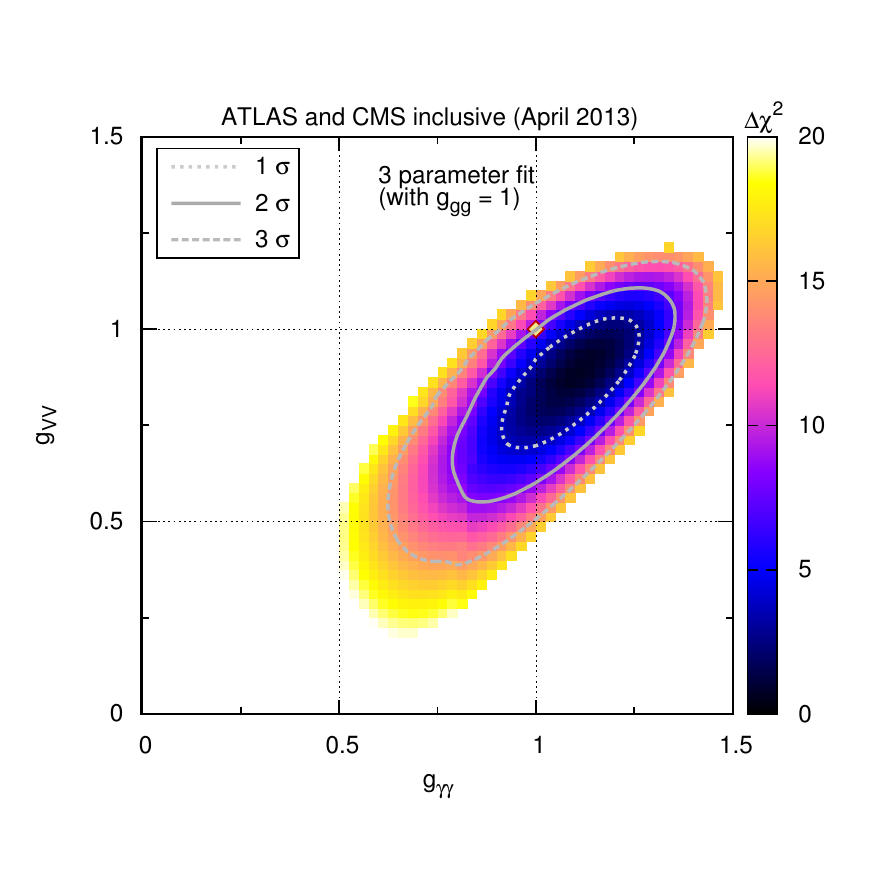}
\centering\includegraphics[scale=.9]{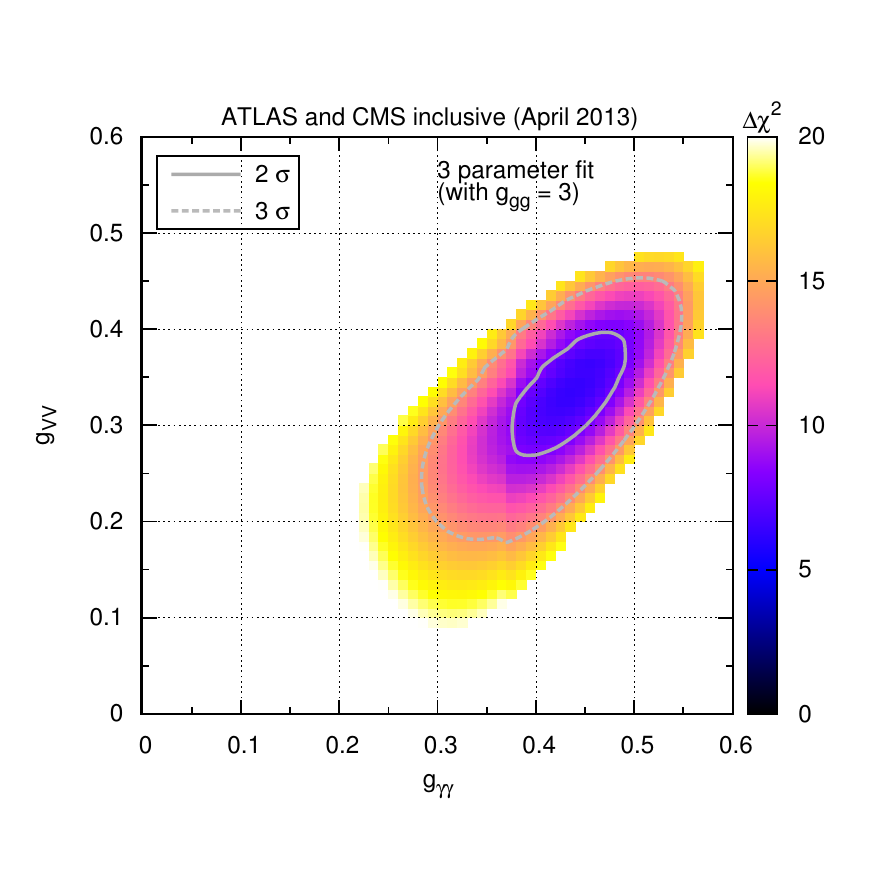}
\vspace{-7ex}
\caption{Same as Fig.~(3) using inclusive $H\rightarrow\gamma\gamma$ data.\hspace{50ex}}\end{figure}
\begin{figure}[]
\vspace{-5ex}
\centering\includegraphics[scale=.9]{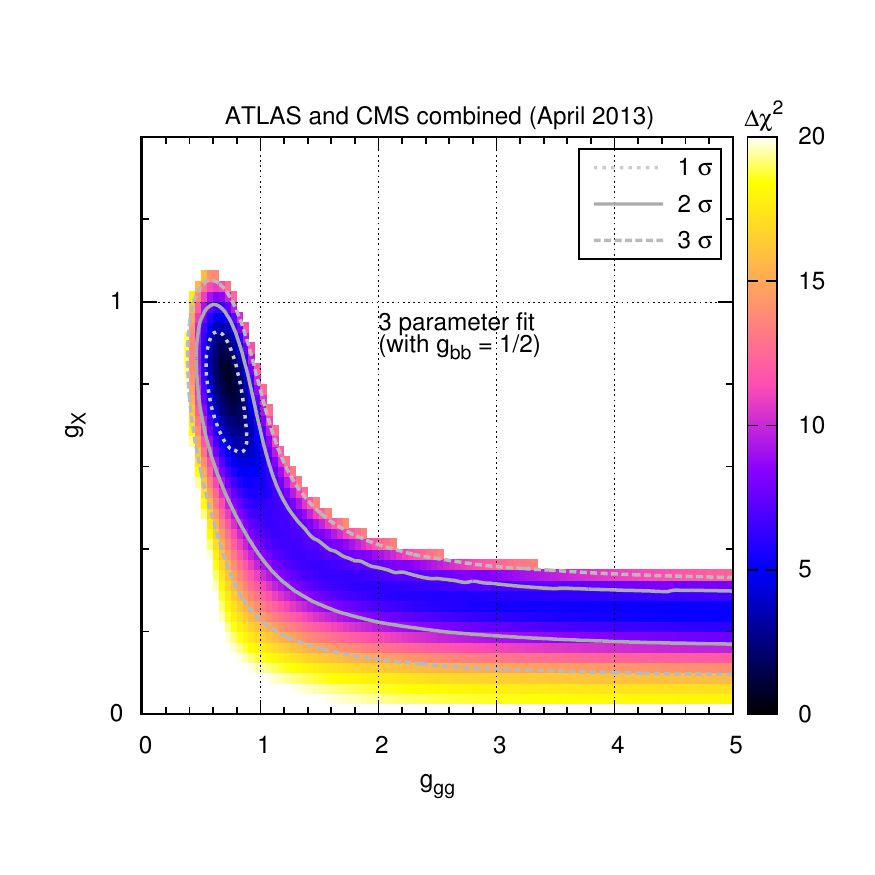}
\centering\includegraphics[scale=.9]{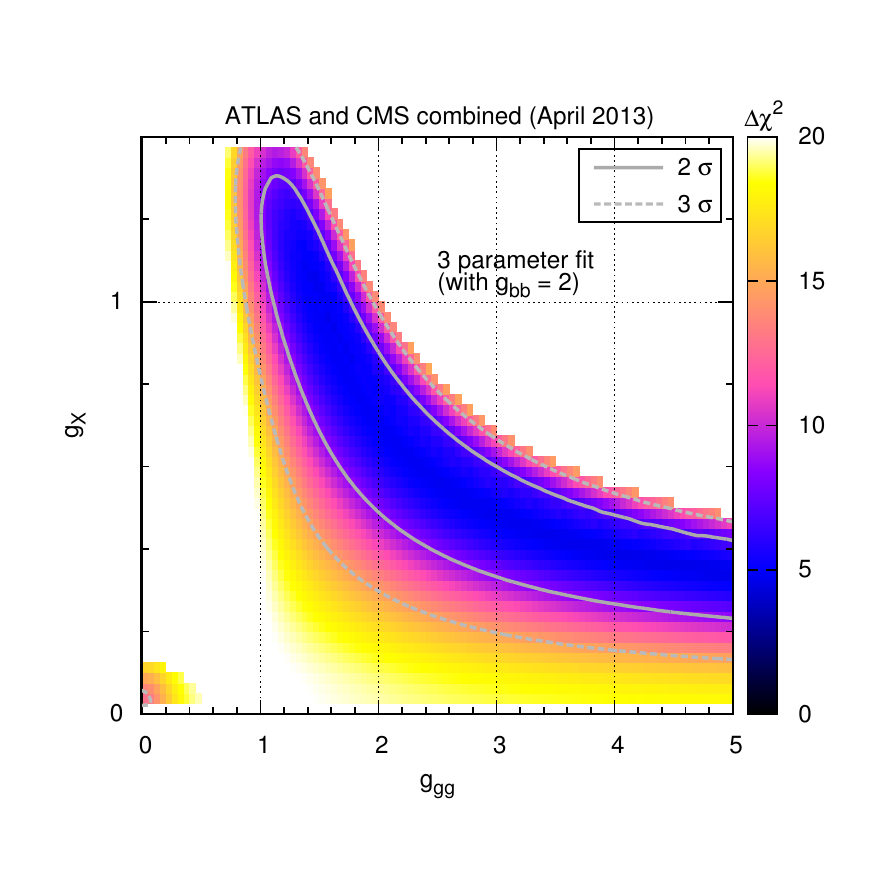}
\vspace{-7ex}
\caption{Three parameter ($g_{bb}$, $g_{gg}$, $g_X$) global fit shown for $g_{bb}=1/2$ (left) and $g_{bb}=2$ (right).}\end{figure}

We can obtain other views of the $\Delta\chi^2$ distribution from the 3 parameter fit by considering the planes defined by fixed $g_{gg}$. Two examples for $g_{gg}=1$ and 3 are shown in Fig.~(3).\footnote{If the Tevatron data is included then there is no longer a region enclosed within a 2 $\sigma$ contour in the $g_{gg}=3$ plot.} For $H\rightarrow\gamma\gamma$ the collaborations present their results in various categories. They also present inclusive results for $H\rightarrow\gamma\gamma$, and HiggsSignals provides the option to use those results instead. We present the analog of Fig.~(3) in this case in Fig.~(4), where the global minimum is at $g_{gg}=0.88$, $g_{\gamma\gamma}=1.22$, $g_{VV}=0.95$ with $\chi^2/{\rm ndf}=18.6/14$. In this case we see that the SM point sits on the 2 $\sigma$ contour.

To explore the effect of changing the $b\overline{b}$ coupling, we may consider another 3 parameter fit with parameters $g_{bb}$, $g_{gg}$ and with $g_X$ defined as before. We return to the categorized $H\rightarrow\gamma\gamma$ data and find the global minimum at $g_{bb}=0.84$, $g_{gg}=0.81$, $g_{X}=0.93$ with $\chi^2/{\rm ndf}=26.3/33$. Here we consider two planes defined by $g_{bb}=1/2$ and $g_{bb}=2$, with the results shown in Fig.~(5).

 These various results indicate that couplings far from the standard model values are not ruled out by the LHC, including $W$ and $Z$ couplings roughly one-third of the SM strength. For such couplings the ``Higgs'' has less to do with EWSB, and this leaves an opening for a more natural origin of EWSB. New colored fermions can be a natural part of such a description, and loops of these fermions will increase the coupling to gluons. Thus the increase in the production rate and the reduced $W$ and $Z$ couplings can go hand in hand.  Note also that constraints on these couplings from electroweak precision tests are also modified by new contributions to the oblique parameters.

An example of a particle with reduced couplings to $W$ and $Z$ is a state that is a mixture of a light pseudoscalar and a heavy scalar. This mixing requires CP violation. The coupling to $W$ and $Z$ is determined by the size of the scalar component of the mixed state. This should not be large since the mixed state is light, which means that the pseudoscalar component, a pseudo-Goldstone boson, should dominate. And unlike the $gg$ coupling which is enhanced, the $\gamma\gamma$ coupling is reduced (the pseudoscalar component lacks the $W$ loop contribution and the scalar component coupling is reduced by the additional loops of heavy fermions). And finally the $\tau\tau$ coupling is reduced if the mixed state is more dominated by heavy quark components vs heavy lepton components.\footnote{Reduced couplings to both $W$'s and electrons reduce the contributions to the electron electric dipole moment.} We thus see how an increased gluon fusion production cross section can occur together with reduced branching ratios. This picture is described in \cite{Holdom:2013pu}. 

The effective description of such a theory is a CP violating multi-Higgs-doublet model. The new ingredient is that the effective couplings of the neutral bosons to gluons and photons are altered by the presence of new heavy chiral fermions. As we have indicated, this opens up a viable region of parameter space in such models where a light state that is mostly pseudoscalar can resemble the presently observed properties of the 126 GeV boson. Note in particular that it is the scalar component that manifests itself in the angular distributions of the four leptons in the $ZZ$ decay mode. Evidence for the pseudoscalar component can be searched for in other processes.

Most importantly, bosons that are scalar or mostly scalar are not required to have unnaturally small masses. We conclude that the present LHC data still leaves open a very non-Higgs-like interpretation of the 126 GeV boson.

\section*{Acknowledgments}
This work was supported in part by the Natural Science and Engineering Research Council of Canada.

\end{document}